\documentclass[12pt]{iopart}
\usepackage{graphicx}

\newcommand{\sqrtsNN}{\mbox{$\sqrt{\mathrm{s}_{_{\mathrm{NN}}}}$}}
\newcommand{\axi}{$\overline{\Xi}^+$}
\newcommand{\xim}{$\Xi^-$}
\newcommand{\alam}{$\overline{\Lambda}$}
\newcommand{\lam}{$\Lambda$}
\newcommand{\ks}{$\mathrm{K}^{0}_{S}$}
\newcommand{\omm}{$\Omega^-$}
\newcommand{\aom}{$\overline{\Omega}^+$}

\begin{document}

\title[$\Xi$ and $\Omega$ v$_{2}$ at \sqrtsNN=200 GeV]{Elliptic flow of multi-strange baryons $\Xi$ and $\Omega$
in Au+Au collisions at \sqrtsNN=200 GeV}

\author{Javier Castillo for the STAR Collaboration
\footnote[1]{For the full author list and acknowledgments, see
Appendix ``Collaborations'' of this volume.}}

\address{Lawrence Berkeley National Laboratory, Berkeley, CA 94720, USA}

\ead{JECastillo@lbl.gov}

\begin{abstract}
The first measurement of the elliptic transverse flow for
multi-strange baryons $\Xi$ and $\Omega$ in high energy heavy ion
collisions is presented, which may indicate the presence of
partonic collectivity. A hydrodynamically inspired model fit to
the transverse momentum spectra and elliptic flow of \xim and \axi
indicates that these particles might be emitted from the system at
a high temperature ($\sim 150$ MeV) with significant radial
transverse flow and that the emitting system is spatially
asymmetric.
\end{abstract}



%\maketitle

\section{Introduction}
In heavy ion collisions we aim to investigate nuclear matter under
extreme conditions of pressure and temperature which is expected
to lead to the creation of deconfined partonic matter, the Quark
Gluon Plasma (QGP). Lattice QCD calculations predict the
transition from this partonic system to a hadronic state at
$T_c\approx150 -- 180$ MeV~\cite{Lqcd}. In this thermalized state
that is the QGP, collective effects among constituents such as
transverse flow will develop. Due to the initial spacial asymmetry
of the system in non central collisions a strong and self
quenching elliptic component of the transverse flow should also be
present. Since transverse flow is cumulative and should not be
affected by the hadronization process, the final observed
transverse flow will have a contribution from the partonic stage.
Multi-strange baryons have been suggested to be sensitive to the
early stage of the collision~\cite{StarMSB130} due to their
predicted low hadronic cross sections~\cite{vanHecke98}. Elliptic
flow, due to its self quenching nature, has also proven to be a
good tool for understanding the properties of the early stage of
the collisions. Thus multi-strange baryon elliptic flow could be a
valuable probe of the initial partonic system.

\section{Multi-strange baryons reconstruction and analysis}

\begin{figure}
\begin{center}
\begin{minipage}[c]{0.50\textwidth}
\begin{center}
\includegraphics[width=\textwidth]{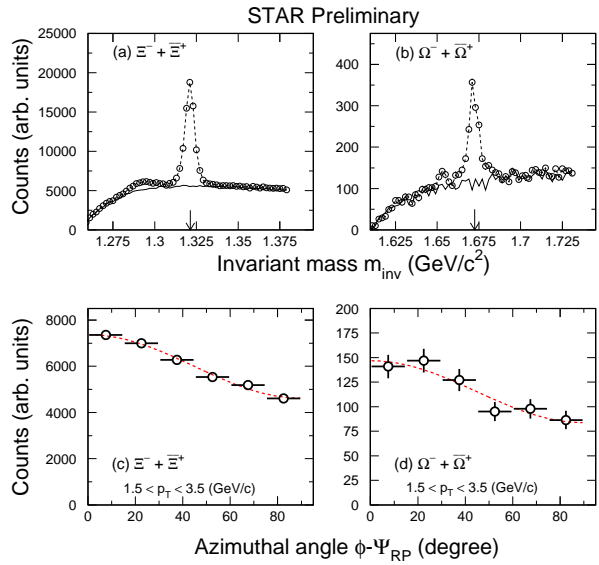}
\end{center}
\end{minipage}\hfill
\begin{minipage}[c]{0.48\textwidth}
\caption{(a) \xim+\axi and (b) \omm+\aom invariant mass
distribution for minimum bias events (0--80\%) from Au+Au
collisions at \sqrtsNN=200 GeV. The solid lines show the remaining
combinatorial background as estimated from a same event rotating
method (see text for details). Azimuthal distribution with respect
to the reaction plane for (c) \xim+\axi and (d) \omm+\aom raw
yields.} \label{Fig:InvMassXiOm}
\end{minipage}
\end{center}
\end{figure}

In this paper we present the first measurement of the azimuthal
anisotropy (characterized by the elliptic flow parameter $v_2$) of
multi-strange baryons $\Xi$ and $\Omega$. Using the STAR detector
at RHIC, multi-strange baryons are reconstructed via the topology
of their decay $\Xi \rightarrow \Lambda+\pi$ and $\Omega
\rightarrow \Lambda+K$ followed by $\Lambda \rightarrow p+\pi$.
Simple cuts on geometry, kinematics and particle identification
via specific ionization are applied to reduce the combinatorial
background. Figure~\ref{Fig:InvMassXiOm} shows the invariant mass
distribution for (a) \xim+\axi and (b) \omm+\aom in minimum-bias
collisions (0-80\% of the total hadronic cross-section) where a
clear peak can be seen over the remaining combinatorial
background. The background can be determined by sampling two
regions on either side of the peak. It can also be reproduced by
rotating the $\Lambda$ candidates by $180^{\circ}$ and then
reconstructing $\Xi$ and $\Omega$ candidates (solid lines in
Fig.~\ref{Fig:InvMassXiOm} (a) and (b)). Both methods give
equivalent results. We calculate $v_{2}$ from the distribution of
the particle raw yields as a function of the azimuthal angle with
respect to the reaction plane. The $\Xi$ and $\Omega$ candidates
are thus divided in $\phi-\Psi_{RP}$ bins, and the raw yields for
each bin are extracted from the invariant mass distributions as
described above. The reaction plane is estimated by the event
plane which is calculated from the azimuthal distribution of
primary tracks. To avoid autocorrelation, tracks associated with a
$\Xi$ or $\Omega$ candidate are excluded from the event plane
calculation. Figure~\ref{Fig:InvMassXiOm} shows the
$\phi-\Psi_{RP}$ raw yield distributions of (c) \xim+\axi and (d)
\omm+\aom from the minimum bias data set in the
$1.5<\mathrm{p}_{\bot}<3.5$ GeV/c range. These distributions
exhibit a clear $\cos(2(\phi-\Psi_{RP}))$ oscillation indicating
that both $\Xi$ and $\Omega$ particles have non zero elliptic
flow. Furthermore, we note that the oscillations for $\Xi$ and
$\Omega$ are of the same magnitude as will be discussed later. The
observed $v_{2}$ is corrected to account for the finite resolution
of the event plane~\cite{RPresolution}. We calculate this event
plane resolution by the random subevents method. Finally, the
corrected transverse momentum spectra are obtained as described
in~\cite{StarMSB130}.

\section{Results and discussions}

\begin{figure}
\begin{minipage}[t]{0.49\textwidth}
\begin{center}
\includegraphics[width=0.98\textwidth]{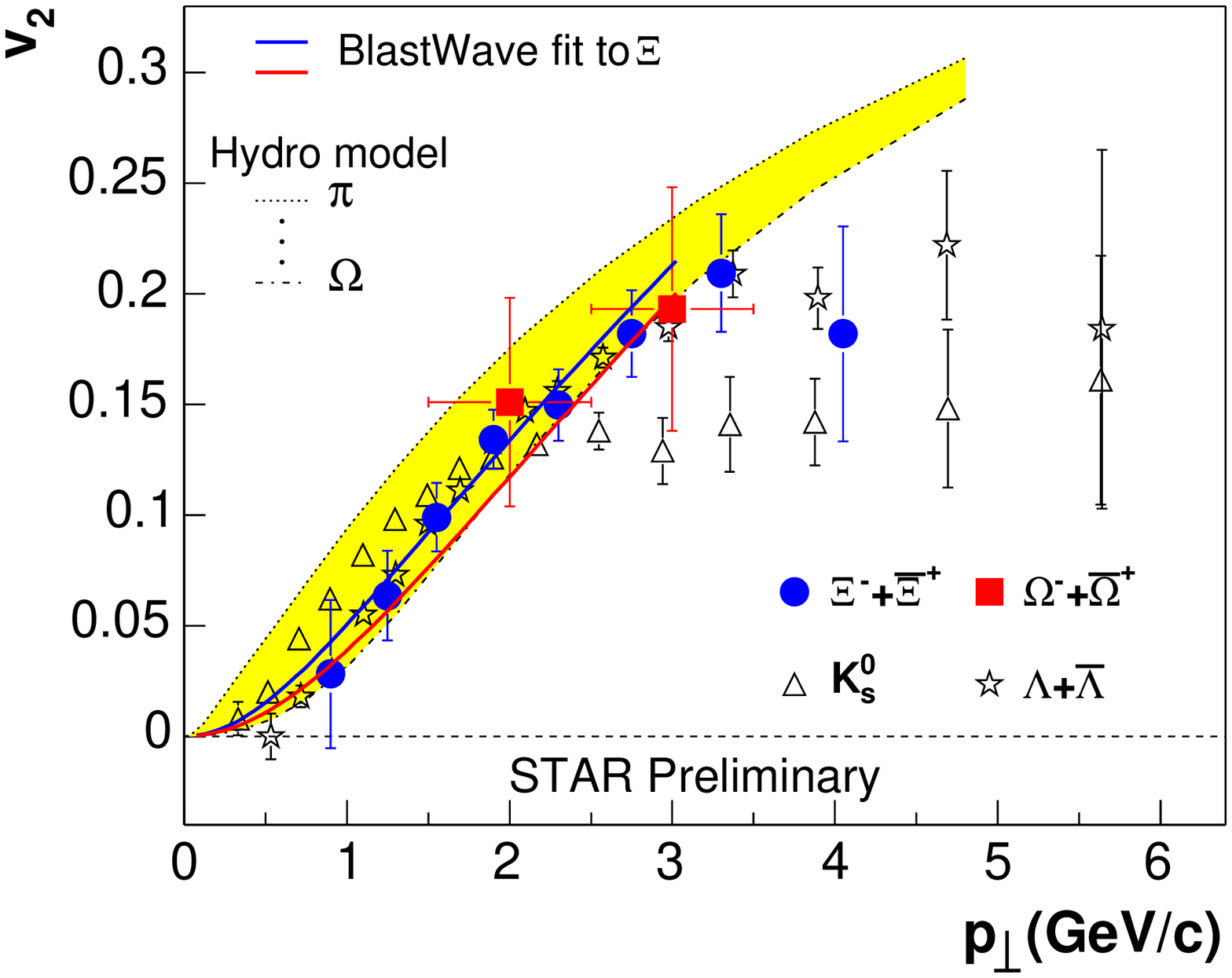}
\caption{$v_2(p_{\bot})$ for \xim+\axi and \omm+\aom from minimum
bias collisions. The results for \ks ~and \lam+\alam ~are also
shown. See text for a description of the curves.}
\label{Fig:V2PtXiOm}
\end{center}
\end{minipage}\hfill
\begin{minipage}[t]{0.49\textwidth}
\begin{center}
\includegraphics[width=0.98\textwidth]{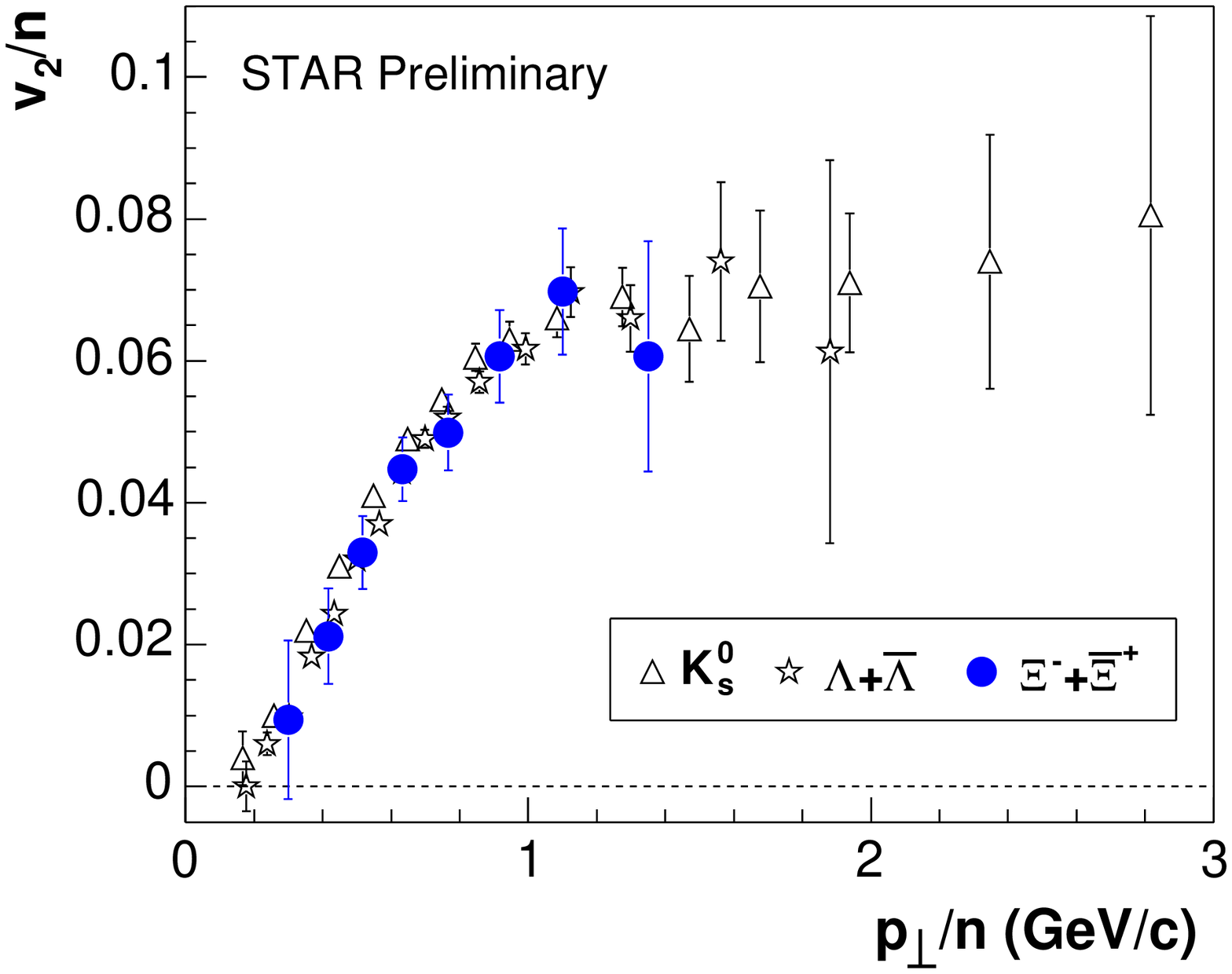}
\caption{$v_{2}/n$ as a function of $p_{\bot}/n$ for \ks,
\lam+\alam~\cite{StarK0sLamV2} and \xim+\axi, where $n$ is the
number of constituent quarks for each particle.}
\label{Fig:QuarkCoal}
\end{center}
\end{minipage}\hfill
\end{figure}

Figure~\ref{Fig:V2PtXiOm} shows the first measurement of $v_2$ for
multi-strange baryons \xim+\axi and \omm+\aom as a function of
$p_{\bot}$ for the minimum bias data set. We first observe that
the $v_2$ for $\Xi$ increases with $p_{\bot}$ reaching a
saturation value of $\sim 0.18$ at $p_{\bot}\sim3.0$ GeV/c.
Secondly, the $v_2(p_{\bot})$ for $\Omega$ in the measured
$p_{\bot}$ range is clearly non-zero and is consistent with the
one measured for $\Xi$, indicating that even the triply-strange
baryon $\Omega$ shows significant elliptic flow. Also shown in
Fig.~\ref{Fig:V2PtXiOm} are the $v_2$ for \ks and
\lam+\alam~\cite{StarK0sLamV2}. The colored band represent typical
hydrodynamic model calculations~\cite{Pasi01} from $\pi$ to
$\Omega$ mass and the results of a hydro-inspired model fit
(discussed later) are shown as solid lines. We observe that the
$p_{\bot}$ dependence of the elliptic flow parameter for $\Xi$ is
similar to that of the \lam. This supports the previously
established baryon to meson dependence of the elliptic flow
parameter~\cite{StarK0sLamV2}. This particle type dependence of
the $v_2(p_{\bot})$ in the intermediate $p_{\bot}$ region is
naturally accounted for by the quark coalescence or recombination
models~\cite{Molnar03,Greco03,Fries03} with the underlying
assumption of partonic collectivity. In such hadronization models,
hadrons are formed by quark coalescence from a partonic system.
The particle $v_2(p_{\bot})$ is then predicted to show number of
constituent quarks scaling. Figure~\ref{Fig:QuarkCoal} shows the
scaled $v_2/n$ versus $p_{\bot}/n$ where $n$ is the number of
constituent quarks of the particle for \ks,
\lam~\cite{StarK0sLamV2} and $\Xi$. The coalescence approach seems
to hold for multi-strange baryons. This could be another
indication of the presence of partonic collectivity at the early
stage of the collision. Furthermore, the partonic flow of the $s$
quark seems to be of similar magnitude than that of the $u$ and
$d$ quarks.

\begin{figure}[tb]
\begin{center}
\includegraphics[width=0.70\textwidth]{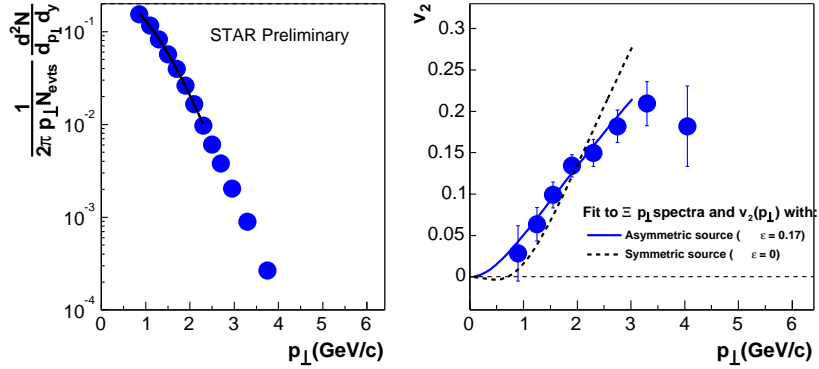}
\caption{The minimum bias (a) $p_{\bot}$ and (b) $v_{2}(p_{\bot})$
of \xim+\axi. The lines are the results from the hydro inspired
model fits to the \xim and \axi $m_{\bot}$ spectra and \xim+\axi
$v_{2}(p_{\bot})$. The solid line corresponds to the best fit
($\varepsilon=0.17\pm0.04$) while the dashed line corresponds to
the fit with fixed eccentricity $\varepsilon=0.0$.}
\label{Fig:qm04XiV2VsPtBW}
\end{center}
\end{figure}

A simultaneous fit of the $p_{\bot}$ spectra and the
$v_{2}(p_{\bot})$ of \xim and \axi for the minimum bias data set
can be performed using an extension~\cite{Retiere03} of a
hydro-inspired model~\cite{Schnedermann93}. In this model, all
considered particles are emitted from a thermal expanding source
with a transverse rapidity $\rho=\tanh^{-1}(\beta)$, where $\beta$
is the transverse velocity, at the thermal freeze-out temperature
$T_{\mathrm{fo}}$. Furthermore, the transverse flow rapidity is
assumed to be asymmetric with respect to the reaction plane, and
can be described as $\rho=\rho_{0}+\rho_{a}\cos(2\phi)$. Finally,
a spatial asymmetry of the source at freeze-out can be introduced
and described by the eccentricity
$\varepsilon=\frac{R^2_y-R^2_x}{R^2_y+R^2_x}$ where $R_x$ and
$R_y$ are the source radii in the in-plane and out-of-plane
directions respectively. Using the implementation described
in~\cite{Retiere03} to fit our data in the range
$0.7<p_{\bot}<2.5$ GeV, we obtain $T_{\mathrm{fo}}=142\pm17$ MeV,
$\rho_{0}=0.80\pm0.05$ ($\langle \beta_{\bot}
\rangle=0.47\pm0.03$), $\rho_{a}=0.047\pm0.017$ and
$\varepsilon=0.17\pm0.05$. As previously
observed~\cite{StarMSB130}, the $\Xi$ data is, within this
framework, best described with high $T_{\mathrm{fo}}$ and low
$\langle \beta_{\bot} \rangle$. We also note that the
$v_2(p_{\bot})$ for $\Xi$ seems to favor an out-of-plane extended
source at freeze-out. Indeed, the solid line in
figure~\ref{Fig:qm04XiV2VsPtBW} shows the $v_2(p_{\bot})$ of $\Xi$
which results from the previous fit, {\it i.e.}
$\varepsilon=0.17$, together with the result of a fit which
requires the source to be azimuthally symmetric, {\it i.e.}
$\varepsilon=0.0$ (dashed line). Clearly the data is best
described by an asymmetric source. We note that at \sqrtsNN=130
GeV, the STAR $v_2$ of $\pi$, $K$ and $p$ also required the
emitting source to be spatially asymmetric and found
$\varepsilon=0.04\pm0.01$ for the minimum bias data set
(0--85\%)~\cite{StarPIDV2}. Also at \sqrtsNN=200 GeV, the source
eccentricity at the freeze-out of the $\pi$ as calculated from
azimuthally sensitive $\pi$ HBT ranges from $\varepsilon\sim0.01$
for the most central collisions to $\varepsilon\sim0.13$ for the
most peripheral bin~\cite{StarPiAzHBT}. Finally, a fit to the
PHENIX $\pi$, $K$ and $p$ spectra and
$v_2$~\cite{PhenixSpectra,PhenixV2} results in
$\varepsilon=0.121\pm0.004$~\cite{RetiereQM04}.

\section{Conclusion}

In summary, we have reported the first measurement of
multi-strange baryons $\Xi$ and $\Omega$ elliptic flow in high
energy nucleus nucleus collisions which may indicate the presence
of partonic collectivity. Both $\Xi$ and $\Omega$ show a
significant elliptic flow which is of the same magnitude as for
other particles {\it e.g.}\ $\Lambda$. The coalescence approach
seems to also describe the multi-strange baryons $v_2$ at
intermediate $p_{\bot}$. A hydro-inspired thermal model
consistently describes both $\Xi$ $m_{\bot}$ spectra and
$v_2(p_{\bot})$ and requires a high temperature and an asymmetric
source at thermal freeze-out.

\end{document}